\documentclass[final]{beamer}
\usepackage[orientation=portrait,size=a0,
            scale=1.25         
           ]{beamerposter}
           
\newcommand{\bldz}[1]{{\bf \footnotesize #1}}
\usepackage[utf8]{inputenc}

\title
[Poster E1.16-0085-21 $\bullet$ 43th COSPAR Scientific Assembly, 28 January -- 4 February 2021, Sydney, Australia] 
{ 
Proton-induced activation cross sections in the energy range below 1 GeV
}

\author{ 
I.~V.~Moskalenko\inst{1}, A.~A.~Andrianov\inst{2}, A.~V.~Bytenko\inst{3}, T.~A.~Frolova\inst{2}, R.~S.~Khalikov\inst{4}, A.~Yu.~Konobeev\inst{5}, Yu.~A.~Korovin\inst{2}, T.~V.~Kulevoy\inst{4}, I.~S.~Kuptsov\inst{2}, K.~V.~Pavlov\inst{4}, A.~Yu.~Stankovskiy\inst{6}, E.~M.~Syresin\inst{3}, Yu.~E.~Titarenko\inst{4}, V.~D.~Vu\inst{2}
}
\institute
[] 
{
\inst{1} HEPL \& KIPAC, Stanford University, Stanford, CA 94305, USA;\ 
\inst{2} Obninsk Institute for Nuclear Power Engineering, NRNU MEPhI, Obninsk 249040, Russia;\ 
\inst{3} Joint Institute for Nuclear Research, Dubna, Moscow Oblast, 141980 Russia;\ 
\inst{4} NRC ``Kurchatov Institute'' -- ITEP, Moscow, 117218 Russia;\ 
\inst{5} Institute for Neutron Physics and Reactor Technology, KIT, 76021 Karlsruhe, Germany;\ 
\inst{6} SCK CEN, Boeretang 200, B-2400 Mol, Belgium
}
\date{\today}

\begin{document}
\begin{frame}[t]
\begin{multicols}{3}

\section{Introduction}

The last decade brought spectacular advances in astrophysics of cosmic rays (CRs) and $\gamma$-ray astronomy. Launches of missions that employ forefront detector technologies were followed by a series of remarkable discoveries even in the energy range that deemed as well-studied. Among those missions are the Payload for Antimatter Matter Exploration and Light-nuclei Astrophysics (PAMELA), the Fermi Large Area Telescope (Fermi-LAT), the Alpha Magnetic Spectrometer-02 (AMS-02), NUCLEON experiment, CALorimetric Electron Telescope (CALET), DArk Matter Particle Explorer mission (DAMPE), and Cosmic-Ray Energetics and Mass investigation (ISS-CREAM). 
Outstanding results have been also delivered by mature missions, such as the Cosmic Ray Isotope Spectrometer onboard of the Advanced Composition Explorer (ACE-CRIS) operating for more than two decades, and by Voyager 1, 2 spacecraft, currently at 151 AU/126 AU from the Sun, respectively. 

The modern technologies employed by many of these missions have enabled measurements with unmatched precision, enabling searches for the subtle signatures of dark matter (DM) and new physics in CR and $\gamma$-ray data. The reached level of precision demonstrates that we are on the verge of major discoveries.

Interpretation of CR data, observations of the diffuse photon emissions, and search of new physics are critically dependent on the accuracy of the isotopic production cross sections. The latter are the centerpiece of any propagation code, but their accuracy has always been the Achilles' heel of propagation models. Handling this long-standing issue requires combined efforts of professional nuclear physicists, both theorists and experimentalists, and astrophysics community. Our goal is a reliable library of nuclear cross section that can be used in many applications.




\section{Current state}

Interactions of nucleons and particles with nuclei in the energy range from several dozens of MeV to several GeV/nucleon are usually described by a combination of Intranuclear cascade (INC), pre-equilibrium, and equilibrium models. More sophisticated Quantum Molecular Dynamics (QMD) models, due to their complexity, did not spread much to really compete with INC models, although they generally show better results at low energies.

The first stage of the reaction assumes that particle-nucleus interactions can be represented by free particle-particle collisions inside the nucleus. Free particle collisions form a sequence often called INC [1]. Nowadays, three variants of INC models having statistical nature are recognized as most developed: Bertini, CEM, INCL.

Bertini [2] considers nucleus as three-region approximation to the continuously changing density distribution of the matter in the nucleus. Nucleons are considered as Fermi gas and the cascade initiated by incident particle is followed either until secondary particles escape the nucleus or when their energy falls below a certain cut-off. Most sophisticated developments of this model are realized in FLUKA [3] and GEANT4 [4] transport codes. In FLUKA, in its nuclear interaction model called PEANUT, the  Generalized INC (GINC) model considers more radial zones, more quantum effects, smooth transition to pre-equilibrium stage. Binary cascade model of GEANT4 takes some advantages of QMD models to propagate particles in nuclear potential. 

The INC Li\`ege (INCL) model [5] is based on classical mechanics, the only quantum effect considered is dynamic Pauli blocking. All collisions are considered as instantaneous at a point because the duration of events is short compared to the time between successive collisions. The most recent version, INCL4.6, has replaced binary cascade as default model choice in GEANT4. 

The Cascade-Exciton Model (CEM) [6] describes the INC using the time-independent Dubna cascade model [7] that makes use of 3D geometry for all cascades.  

All the INC models are followed by pre-equilibrium and equilibrium models, describing intermediate and slow processes of nucleus de-excitation, respectively. However, INCL4.6 does not consider pre-equilibrium stage and is directly coupled to equilibrium models GEM [8] in PHITS transport code [9] or ABLA [10] in GEANT4.

PEANUT of FLUKA and CEM consider their own sequences of models. PEANUT is using modified hybrid exciton model of Blann [11] for pre-equilibrium part and Weisskopf-Eving statistical model [12] for evaporation. For light nuclei Fermi break-up model is implemented. In CEM, Modified Exciton Model [13] and GEM equilibrium model are employed, Fermi break-up is also considered.


\vskip1ex
\begin{figure}[t]
\hskip-1ex\includegraphics[width=0.7\columnwidth]{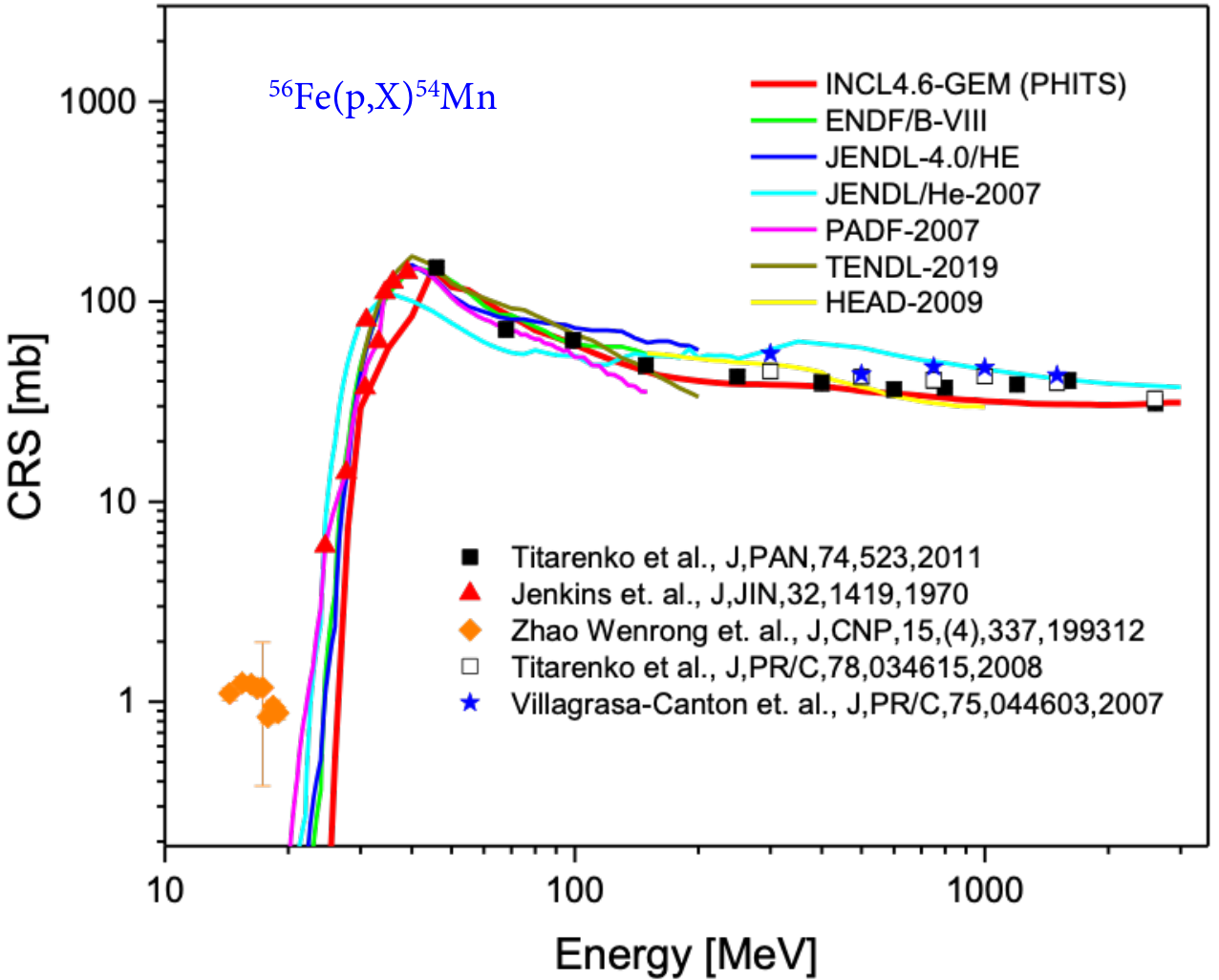}\\
\includegraphics[width=0.68\columnwidth]{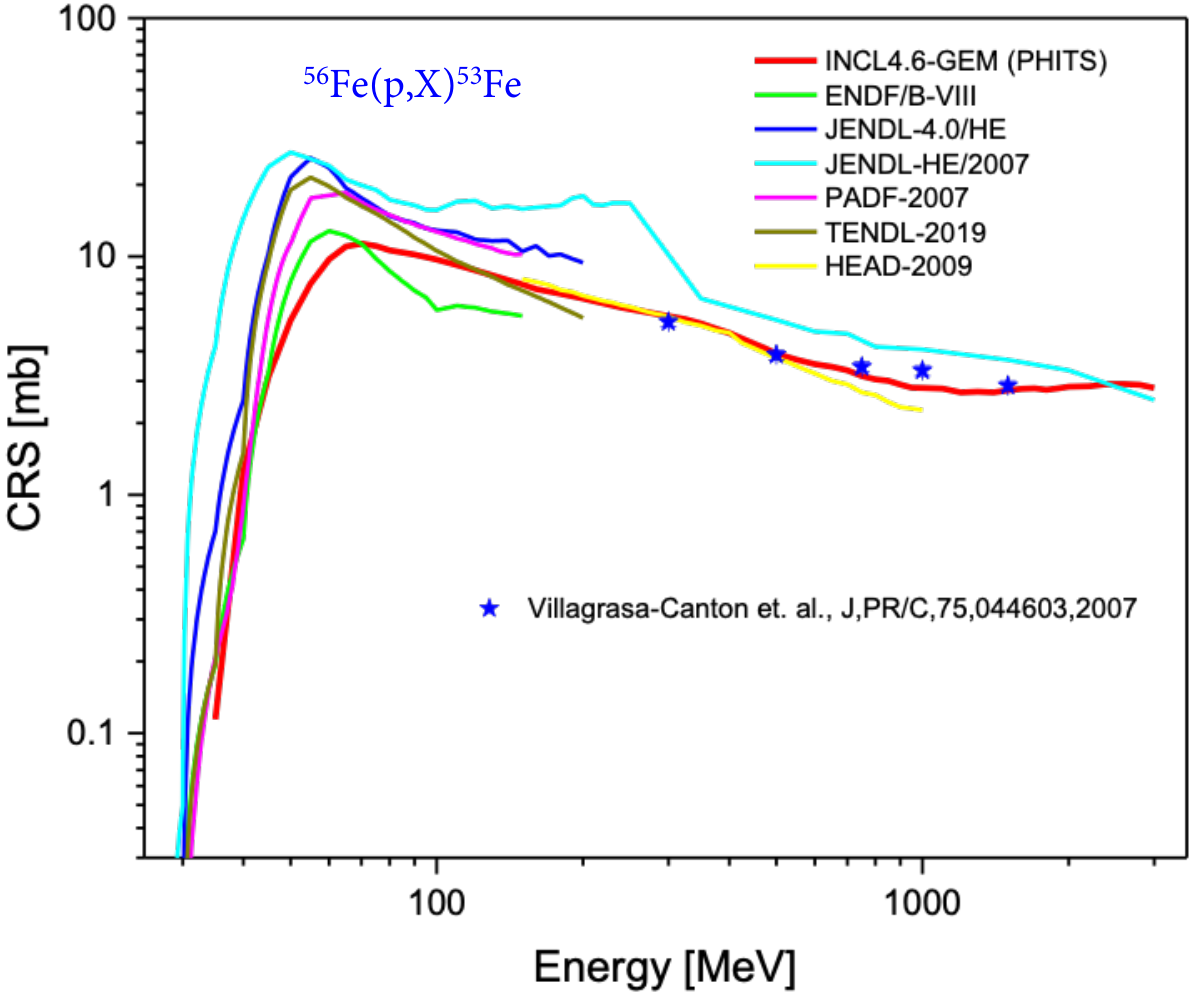}\\
\includegraphics[width=0.68\columnwidth]{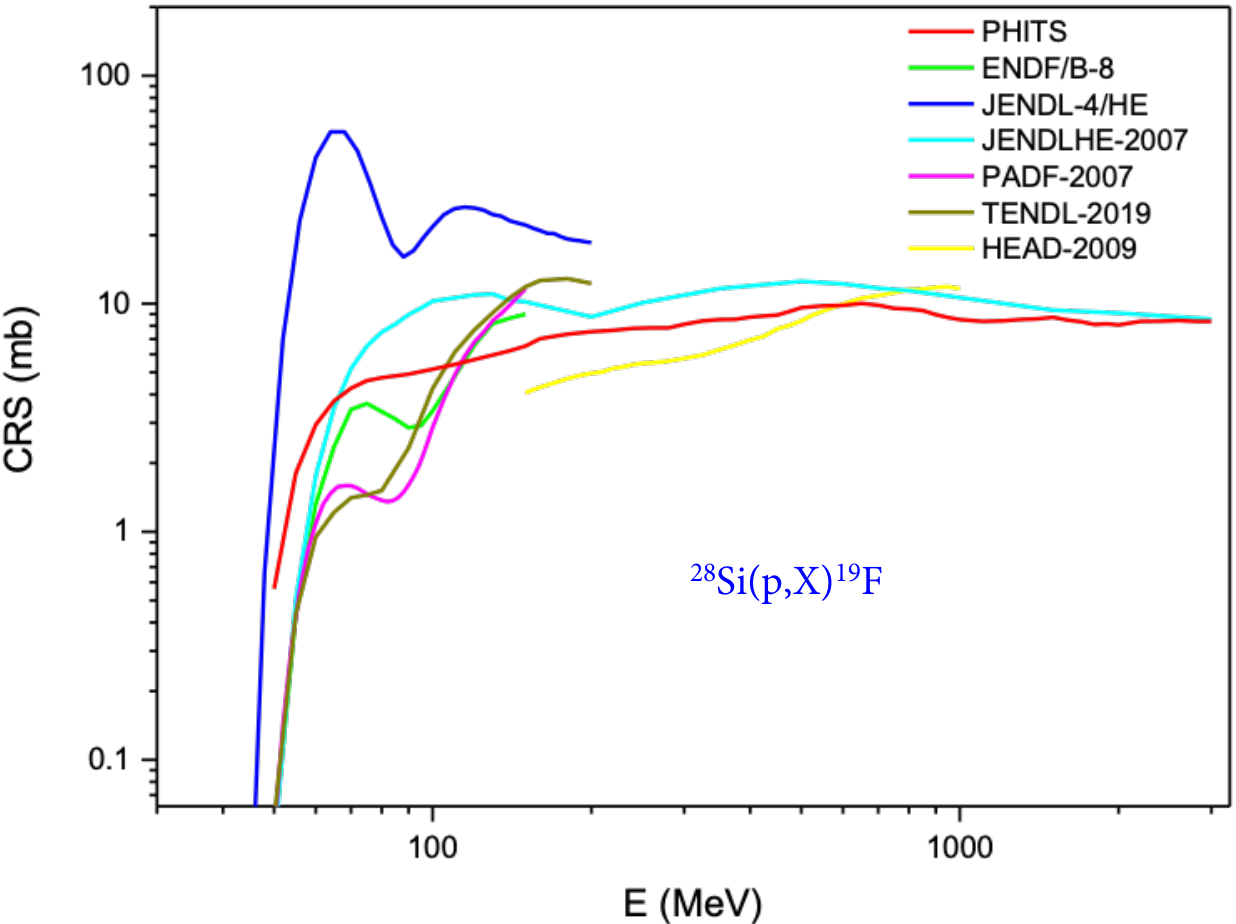}
\caption{\small Example cross sections for reactions $^{56}$Fe($p,X$)$^{54}$Mn (top), $^{56}$Fe($p,X$)$^{53}$Fe (middle), and $^{28}$Si($p,X$)$^{19}$F (bottom). The two top plots show individual cross sections calculated using various nuclear codes and available data. The bottom plot with $^{28}$Si target shows cumulative cross section with contributions from the decay chains originating in secondary nuclides: $^{19}$C, $^{19,20}$N, $^{19}$O, $^{19}$Ne. In the absence of data, calculations show an order of magnitude discrepancies at low energies, which translates into large uncertainties in the interpretation of astrophysical data.}
\label{f1}
\end{figure}
\vskip0ex

A restriction for all INC models at low energies is that De Broglie wavelength of particles is becoming larger than characteristic nuclear size. Despite their significant improvement at $\le$100 MeV/nucleon, deterministic models are generally recommended for this energy range.    

Most complete description of nuclear reactions in the range of 10s MeV is provided by TALYS [14], although it has constraints for atomic masses $<$10 and for energies $<$200 MeV. 
A comparison of evaporation mechanisms manifests in favor of Hauser-Feshbach approach in the intermediate energy range $<$200 MeV/nucleon. This led to attempts of combining INC stage with Hauser-Feshbach statistical emission. Such codes as CASCADEX (combining CASCADE/INPE realizing Dubna INC model with TALYS 1.0) [18], cascadeXO (CASCADE/INPE + TALYS 1.95), cemcas (CEM03.03 + TALYS 1.95) demonstrate better accuracy for certain reactions. 
However, there is no unique model that can describe the cross section data for all nuclei up to several GeV/nucleon. Therefore, it is reasonable to represent the excitation functions as best estimate value accompanied by an uncertainty band. 

\vskip1ex
\begin{table}
\small
\centering
\caption{Summary of INC models}
\begin{tabular}{ccccc}
\hline\hline
& Pre-equi- & De-exci- & Validity & Transport\\
INC & librium & tation & range & code\\
\hline
CEM03 & own & GEM & $\le$5 GeV & MCNP6\\
INCL4.6 & --- & GEM & $\le$3 GeV & PHITS\\
Bertini/INCL++ & own & GEM/ABLA & $\le$10 GeV & GEANT4\\
PEANUT & own & own & $\le$5 GeV & FLUKA\\
\hline\hline
\end{tabular}
\end{table}
\vskip1ex


\section{New Facility}

The major obstacle in improving an accuracy of the nuclear codes is the scarcity of the available measurements (Fig.~\ref{f1}). We are aiming at using the NICA facility (Nuclotron-based Ion Collider fAcility, https://nica.jinr.ru/) -- a new accelerator complex at the Joint Institute for Nuclear Research (JINR, Dubna, Russia). NICA will provide variety of beam species from $p$ to Au ions. Heavy ions will be accelerated up to kinetic energy of 4.5 GeV/nucleon, protons -- up to 12.6 GeV. 
Measurements of the isotopic cross sections will be conducted using proton and ion beams and thin targets. Besides, NICA can be used for studies of secondary neutron and $\gamma$-ray emission -- of interest for simulations of the radiation environment at the rocky surfaces of planets and asteroids. New data will be used to tune models of nuclear reactions and for astrophysical applications.



\subsection{References}
\footnotesize

\noindent
\bldz{[1]} D. Filges \& F. Goldenbaum, Handbook of Spallation Research, Wiley-VCH Verlag GmbH \& Co (2009);
\bldz{[2]} H. W. Bertini, Low-Energy Intranuclear Cascade Calculation, Phys. Rev. 131 (1963) 1801;
\bldz{[3]} G. Battistoni et al., Overview of the FLUKA code, Ann. Nucl. En. 82 (2015) 10;
\bldz{[4]} J. Allison et al., Recent developments in GEANT4, NIMA 835 (2016) 186;
\bldz{[5]} A. Boudard et al., New potentialities of the Li\`ege intranuclear cascade model for reactions induced by nucleons and light charged particles, PRC 87 (2013) 014606;
\bldz{[6]} S. G. Mashnik, A. J. Sierk, CEM03.03 User Manual, LANL report LA-UR-12-01364, 2012;
\bldz{[7]} V. S. Barashenkov and V. D. Toneev, Interaction of High Energy Particle and Nuclei with Atomic Nuclei, Atomizdat, Moscow, 1972;
\bldz{[8]} S. Furihata, Statistical Analysis of Light Fragment Production from Medium Energy Proton-Induced Reactions, NIMB 171 (2000) 252;
\bldz{[9]} T. Sato, et al., Features of Particle and Heavy Ion Transport Code System PHITS Version 3.02, J. Nucl. Sci. \& Technology, 55 (6) (2018) 684;
\bldz{[10]} A. Kelic, et al., Proc. Joint ICTP-IAEA Adv. Workshop on Model Codes for Spallation Reactions, ICTP Trieste, Italy, 4-8 February 2008, IAEA INDC(NDS)-530 (Vienna 2008, p181);
\bldz{[11]} M. Blann, Hybrid model for pre-equilibrium decay in nuclear reactions. PRL 27 (1971) 337;
\bldz{[12]} V. E. Weisskopf, D. H. Ewing, On the Yield of Nuclear Reactions with Heavy Elements, Phys. Rev. 57 (1940) 472;
\bldz{[13]} K. K. Gudima, G. A. Ososkov and V. D. Toneev, Model for the Pre-Equilibrium Decay of Excited Nuclei, Sov. J. Nucl. Phys. 21 (1975) 138;
\bldz{[14]} A. J. Koning, D. Rochman, Modern nuclear data evaluation with the TALYS code system, Nucl. Data Sheets, 113 (2012) 2841;
\bldz{[15]} J. Raynal, ``Notes on ECIS94,'' tech. rep., CEA Saclay Report CEA-N-2772, 1994;
\bldz{[16]} A. Koning and M. Duijvestijn, A global pre-equilibrium model from 7 to 200 MeV based on the optical model potential, NPA 744 (2004) 15;
\bldz{[17]} W. Hauser, H. Feshbach, The Inelastic Scattering of Neutrons, Phys. Rev. 87 (1952) 366;
\bldz{[18]} A. Stankovskiy and A. Konobeyev, CASCADEX -- a combination of intranuclear cascade model from CASCADE/INPE with the Hauser-Feshbach evaporation/fission calculations from TALYS, NIMA 594 (2008) 420.

\vskip1ex
\noindent
{\it \footnotesize Acknowledgements:} IVM acknowledges support from NASA grant No.~NNX17AB48G.






\end{multicols}
\end{frame}
\end{document}